\documentclass[12pt]{article}
\usepackage{latexsym,amsfonts}

\renewcommand{\title}[1]{\begin{center}\bf\Large #1\end{center}}

\renewcommand{\author}[1]{\begin{center}\large #1\end{center}}

\newcommand{\address}[2]{\begin{center}{\it\normalsize
    #1}\\{\rm\normalsize E-mail: #2}\end{center}}

\newcommand{\ppNumber}[1]{\vspace*{-25mm}%
\hspace*{\textwidth}\makebox[0cm][r]{%
\raisebox{5mm}[0mm][0mm]{\parbox{5cm}{\raggedleft\bf #1}}}}

\newlength{\parindentsave}
\newenvironment{frontmatter}{
  \setlength{\parindentsave}{\parindent}\parindent0cm}{
  \parindent\parindentsave}

\newenvironment{keyword}{\begin{quote}{\it Key words:}}{\end{quote}}

\newcommand{\Cset}{{\mathbb C}}
\newcommand{\Rset}{{\mathbb R}}
\newcommand{\Zset}{{\mathbb Z}}
\newcommand{\sll}{SL(2,$\Rset$)}
\newcommand{\slu}{SL(2,$\Rset$)/U(1)}
\newcommand{\glc}{GL(2,$\Cset$)}

\renewcommand{\i}{{\rm i}}
\newcommand{\e}{{\rm e}}
\renewcommand{\d}{{\rm d}}
\newcommand{\const}{{\rm const}}
\newcommand{\halb}{\frac{1}{2}}
\newcommand{\gasq}{{\gamma^2}}
\newcommand{\gasqhalb}{{\frac{\gamma^2}{2}}}
\newcommand{\gasqapi}{{\frac{\gamma^2}{8\pi}}}
\newcommand{\gasqvpi}{{\frac{\gamma^2}{4\pi}}}
\newcommand{\gasqzpisq}{{\frac{\gamma^2}{32\pi^2}}}

\newcommand{\thr}[1]{\tanh^{#1}\!r\;}
\newcommand{\shr}[1]{\sinh^{#1}\!{r}\;}
\newcommand{\chr}[1]{\cosh^{#1}\!{r}\;}
\newcommand{\tr}{{\rm tr}}
\newcommand{\Id}{{1\hspace{-3pt}{\rm I}}_}
\newcommand{\intl}{\int\limits}
\newcommand{\intk}{\intl_0^{2\pi}}% Integration auf dem Kreis
\newcommand{\inv}{^{-1}}% invers
\newcommand{\comm}[2]{\left[#1,#2\right]}
\newcommand{\pk}[2]{\left\{#1,#2\right\}}
\def\matrix#1#2{\left(\begin{array}{#1}#2\end{array}\right)}

\newcommand{\del}{\partial}
\newcommand{\delz}{\partial_z}
\newcommand{\delzb}{\partial_\zb}

\newcommand{\mb}{{\bar{m}}}
\newcommand{\ub}{{\bar{u}}}
\newcommand{\yb}{{\bar{y}}}
\newcommand{\zb}{{\bar{z}}}
\newcommand{\ys}{{\tilde y}}
\newcommand{\ybs}{{\tilde{\bar y}}}
\newcommand{\tp}{\dot{t}}
\newcommand{\dey}{{\delta y}}

\newcommand{\Ab}{{\bar{A}}}
\newcommand{\Bb}{{\bar{B}}}
\newcommand{\Cb}{{\bar{C}}}
\newcommand{\Db}{{\bar{D}}}
\newcommand{\Eb}{{\bar{E}}}
\newcommand{\Mb}{{\bar{M}}}
\newcommand{\Qb}{{\bar{Q}}}
\newcommand{\Vb}{{\bar{V}}}
\newcommand{\Wb}{{\bar{W}}}
\newcommand{\deC}{{\delta C}}

\newcommand{\alphab}{{\bar{\alpha}}}
\newcommand{\betab}{{\bar{\beta}}}
\newcommand{\nub}{{\bar{\nu}}}
\newcommand{\kappab}{{\bar{\kappa}}}
\newcommand{\etab}{{\bar{\eta}}}
\newcommand{\psib}{{\bar{\psi}}}
\newcommand{\Phib}{{\bar{\Phi}}}
\newcommand{\phib}{{\bar{\phi}}}
\newcommand{\chib}{{\bar{\chi}}}
\newcommand{\alphap}{{\alpha'}}%prime
\newcommand{\alphabp}{{\alphab'}}
\newcommand{\deltaper}{\delta_{2\pi}}
\newcommand{\epsper}{\epsilon_{2\pi}}

\begin{document}
\begin{frontmatter}
  \ppNumber{MZ--TH/99--26\\hep-th/9907057}
  \title{Integration of the \slu{} Gauged WZNW Model with Periodic
    Boundary Conditions}
  \author{Uwe M\"uller}
  \address{Institut f\"ur Physik, Johannes-Gutenberg-Universit\"at,\\
    Staudinger-Weg 7, D-55099 Mainz, Germany}{umueller@thep.uni-mainz.de}
  \author{Gerhard Weigt}
  \address{Deutsches Elektronen-Synchrotron DESY Zeuthen,\\
    Platanenallee 6, D-15738 Zeuthen, Germany}{weigt@ifh.de}

\begin{abstract}
  Gauged WZNW models are integrable conformal field theories. We
  integrate the classical \slu{} theory with periodic boundary
  conditions, which describes closed strings moving in a curved
  target-space geometry. We calculate its Poisson bracket structure by
  solving an initial state problem. The results differ from previous
  field-theoretic calculations due to zero modes. For a future exact
  canonical quantization the physical fields are (non-locally)
  transformed onto canonical free fields.
\end{abstract}

\end{frontmatter}

\section{Introduction}
The \slu{} gauged WZNW model has attracted much interest in the past
\cite{BCR,BN,Witten,DVV,BS,Tseytlin,GS,Bilal} before it was
recognized that this non-linear theory is classically integrable
\cite{MuWe}. More generally, we could prove that integrability holds
for any gauged WZNW theory \cite{cmp}.  This was known for nilpotent
gauging only, which yields Toda theories \cite{Balog,GS}. So far we
have completely integrated the non-linear equations of motion of the
classical \slu{} model for a field theoretic case with asymptotic
boundary conditions \cite{cmp}.

In this paper we solve the classical \slu{} theory for periodic
boundary conditions. As a conformal field theory this model describes
a closed string moving in the background of a black hole target-space
metric \cite{Witten}. It is especially interesting for quantization;
quantum mechanical deformations of its metric and a correlated
dilaton \cite{Callan} were obtained in some perturbative manner
\cite{DVV,BS,Tseytlin}. However, these calculations were based on an
incomplete effective action \cite{Buscher,Mu,cmp}.

Our intention is to provide a different understanding of such quantum
mechanical results. Starting with an entirely classical approach
\cite{BCR,cmp}, we expect that the exact classical solution of this
theory will facilitate its exact canonical quantization.  We
calculate, as in the field theoretic case, the Poisson bracket
structure of the theory by solving an initial state problem and look
for a canonical transformation of the physical fields onto canonical
free fields.  But the results for periodic boundary conditions cannot
simply be inferred from the field theoretic ones because additional
zero modes become important.

To make this paper self-contained, we mention in section $2$ some of
our earlier results; more details are found in ref.\ \cite{cmp}.
First we define the theory, give its general solution and inspect the
symmetry properties. Section $3$ solves an initial state problem which
allows us to calculate in section $4$ the Poisson bracket structure of
the model. A free-field realization of these brackets is given in
section 5. The summary provides a canonical transformation of the
physical fields onto the free fields. Some technical details are found
in two appendices.

\section{The \slu{} theory}

The exact action of the \slu{} gauged WZNW theory written in
light-cone coordinates $z=\tau+\sigma$, $\zb=\tau-\sigma$
\begin{equation}\label{action}
S[r,t]=\frac{1}{\gamma^2}\int_{M}\left(
\delz r\delzb r+\thr{2}\delz t\delzb t\right)
\d z\,\d\zb
\end{equation}
was derived entirely classically and in a gauge invariant manner
\cite{BCR,cmp}. $M$ has cylindrical topology where the space-like submanifolds are
topologically equivalent to a circle
\begin{equation}\label{int-bereich-kreis}
M=\Rset\times S^1,\quad\mbox{i.e.}
\quad 0\le\sigma\le 2\pi,\quad -\infty<\tau<\infty.
\end{equation}
The physical fields $r(\sigma,\tau)$, $t(\sigma,\tau)$, which
represent the position of a closed bosonic string in the target-space
at proper time $\tau$, are subject to the boundary conditions
\begin{equation}\label{period-rand-rt}
r(\sigma+2\pi,\tau)=r(\sigma,\tau),\quad
t(\sigma+2\pi,\tau)=t(\sigma,\tau)+2\pi w,\quad w\in\Zset.
\end{equation}
The $t$ coordinate is an angular variable given modulo $2\pi$ only,
and the winding number $w$ tells us how often the string surrounds the
coordinate origin. The string moves in the curved metric of a
Euclidean black hole \cite{Witten}
\begin{equation}\label{targetmetrik}
\d s^2=\d r^2+\thr{2}\d t^2,
\end{equation}
and the dynamics is given by the equations of motion 
\begin{eqnarray}\label{eom}
\hspace{2.5cm}
\delz\delzb r&=&\frac{\shr{}}{\chr{3}}\delz t\delzb t,\nonumber\\
\delz\delzb t&=&-\frac{1}{\shr{}\chr{}}
\left(\delz r\delzb t+\delz t\delzb r\right).
\end{eqnarray}
These equations are integrable because they have a Lax pair
representation
\begin{equation}
  \label{laxcomm}
  \comm{\delzb-\Cb}{\delz-C}=
  \delz \Cb-\delzb C-\comm{C}{\Cb}=0.
\end{equation}
$C$ and $\Cb$ take values in the Lie algebra of the group SL(2,$\Rset$)
\cite{cmp} 
\begin{equation}
  \label{komp1}%Komponenten
  C=C_a T^a,\quad \Cb=\Cb_a T^a,\quad(a=1,2,3)
\end{equation}
with
\begin{equation}
T^1=\matrix{rr}{0&\quad1\\\hspace{-1ex}-1&0},\quad
T^2=\matrix{rr}{1&0\\\hspace{-1ex}0&\;\;-1},\quad
T^3=T^1T^2=\matrix{rr}{0&\;-1\\\hspace{-1ex}-1&0}.
\end{equation}
One can check that
\begin{eqnarray}
  \label{komp2}%Komponenten
  &&C_1=-\halb\thr{2}\delz t,\quad C_2=C_3=0,\quad
  \Cb_1=\halb\thr{2}\delzb t,\nonumber\\
  &&\Cb_2=-\frac{1}{\cosh r}\delzb\left(\sinh r\cos t\right),\quad
  \Cb_3=\frac{1}{\cosh r}\delzb\left(\sinh r\sin t\right)
\end{eqnarray}
makes the flatness condition (\ref{laxcomm}) equivalent to the
equations of motion (\ref{eom}). But unlike Toda theories \cite{LS}
there is at present no general method to integrate a Lax pair
following from a non-nilpotent gauged WZNW model like (\ref{laxcomm})
directly.  We found the general solution of (\ref{eom}) in \cite{MuWe}
by analysing non-abelian Toda theories \cite{GS,Bilal} as
\begin{equation}\label{solution}
\shr{2}=X\bar{X},\quad
t=\i(B-\bar{B})+\frac{\i}{2}\ln\frac{X}{\bar{X}}
\end{equation}
with the definitions
\begin{equation}\label{X-def}
X=A+\frac{{\bar B}'}{{\bar A}'}(1+A\bar{A}),\quad
\bar{X}=\bar{A}+\frac{B'}{A'}(1+A\bar{A}).
\end{equation}
$A=A(z)$, $B=B(z)$, $\bar{A}=\bar A(\zb)$ and $\bar{B}=\bar B(\zb)$
are complex (anti-) chiral parameter functions and $A'(z)$ etc.
derivatives.  However, as we shall see we must restrict this solution
in order to render $r$ and $t$ real. Straightforward substitution
shows that the solution (\ref{solution}) fulfills the equations of
motion (\ref{eom}). But it will become obvious from the investigation
of initial-value problems of section \ref{sect-initial-value} that the
solution (\ref{solution}) exhausts the entire solution space
(excluding singular solutions). 

The theory is also characterized by conservation laws. The equations
of motion (\ref{eom}) guarantee, in particular, conservation and
chirality of the energy-momentum tensor (we shall omit the anti-chiral
parts whenever possible)
\begin{eqnarray}\label{eit}
T\equiv T_{zz}=\frac{1}{\gamma^2}\left(
\left(\delz r\right)^2+\thr{2}\left(\delz t\right)^2\right),&\quad&
T_{z\zb}=0,
\end{eqnarray}
and in addition of parafermionic observables
\cite{FZ,BCR,MuWe}
\begin{equation}\label{vpm}
V_{\pm}=\frac{1}{\gamma^2}\e^{\pm\i\nu}
\left(\delz r\pm\i\thr{}\delz t\right),
\end{equation}
where $\nu$ is defined by
\begin{eqnarray}\label{nue}
  \delz\nu=\left(1+\thr{2}\right)\delz t,&\quad&
  \delzb\nu=\left(1-\thr{2}\right)\delzb t.
\end{eqnarray}
Since the integrability condition of these equations corresponds to
one of the equations of motion (\ref{eom}), the general solution
(\ref{solution}) integrates eqs.\ (\ref{nue}) to
\begin{eqnarray}\label{nue-explizit}
\nu&=&t+\i(B+\bar B)+\i\ln(1+A\bar A)-\frac{\i}{2}\ln(1+X\bar X)+\nu_0.
\end{eqnarray}

The main purpose of this paper is to calculate the Poisson bracket
structure of the model, assuming canonical Poisson brackets for the
physical fields $r$,$t$ and their conjugate momenta. Therefore, we
have to find $A$, $B$, $\bar A$ and $\bar B$ as functions of these
physical variables. In principle, they are given by solving an initial
state problem defined by a second order differential equation of the
Gelfand-Dikii type
\begin{equation}\label{chardgl}
  y''-(\delz V_-/V_-)y'-\gamma^2Ty=0,
\end{equation}
because its two independent solutions $y_1$, $y_2$ are related to
our parameter functions
\begin{equation}\label{y12def}
y_1=\e^B,\quad\quad y_2=A\e^B,
\end{equation}
and the coefficients of (\ref{chardgl}) are functions of $r$, $t$ and
their derivatives. This differential equation simply follows from the
conserved quantities (\ref{eit}, \ref{vpm}) and (\ref{y12def}) as an
identity.

But the functions $A$, $B$, $\bar A$ and $\bar B$ are not
uniquely determined by this procedure. Because the solution
(\ref{solution}) is invariant under the \glc{} transformations
\begin{eqnarray}\label{Loes-Invarianz}
A&\to&T[A]=\frac{aA-b}{cA+d},\nonumber\\
B&\to&T[B]=B+\ln(c A+d),\nonumber\\
\bar A&\to&T[\bar A]=\frac{d\bar A-c}{b\bar A+a},\nonumber\\
\bar B&\to&T[\bar B]=\bar B+\ln(b\bar A+a),\\
&&\left(
\begin{array}[]{cc}
a&-b\\
c&d
\end{array}\right)
\in\mbox{\glc},\nonumber
\end{eqnarray}
they are only given by the physical fields up to four complex
constants. We shall fix this arbitrariness in the next section.

The monodromy properties of the functions $A$, $B$, $\Ab$, $\Bb$ are, 
as well, determined by \glc{} transformations
\begin{eqnarray}\label{period-rand-AB}
A(z+2\pi)&=&T'[A(z)]=\frac{pA(z)-q}{rA(z)+s},\nonumber\\
B(z+2\pi)&=&T'[B(z)]=B(z)+\ln(rA(z)+s),\nonumber\\
\bar A(\zb-2\pi)&=&T'[\Ab(\zb)]=\frac{s\Ab(\zb)-r}{q\Ab(\zb)+p},\nonumber\\
\bar B(\zb-2\pi)&=&T'[\Bb(\zb)]=\Bb(\zb)+\ln(q\Ab(\zb)+p),\\
&&\matrix{cc}{p&-q\\r&s}\in\mbox{\glc}.\nonumber
\end{eqnarray}
We should remark here that the \glc{} transformations act, indeed, in
two different manners. We also find that $\nu$ is not periodic
modulo $2\pi$
\begin{eqnarray}
  \label{nue-period-explizit}
  \nu(\sigma+2\pi,\tau)&=&\nu(\sigma,\tau)+2\pi w+\i\ln(ps+qr).
\end{eqnarray}
Therefore, the conserved quantities $V_\pm$ and $\bar V_\pm$ are not
periodically defined. We can describe their periodicity behaviour
by the conserved total momentum of the field $t$
\begin{equation}\label{deltamue-def}
   P_t=\frac{1}{\gasq}\intl_0^{2\pi}\thr{2}\tp\;\d\sigma'=
   \intl_0^{2\pi}\pi_t(\sigma',\tau)\;\d\sigma'.
\end{equation}
Using the eqs.\ (\ref{nue}), instead of (\ref{nue-period-explizit}) we
obtain for $\nu$ and $\nub$ the periodicity relations in terms of $P_t$
\begin{eqnarray}\label{nue-period}
\nu(\sigma+2\pi,\tau)-\nu(\sigma,\tau)&=&
\intl_\sigma^{\sigma+2\pi}\nu'(\sigma',\tau)\;\d\sigma'=
2\pi w+\gasq P_t,\nonumber\\
\nub(\sigma+2\pi,\tau)-\nub(\sigma,\tau)&=&
\intl_\sigma^{\sigma+2\pi}\nub'(\sigma',\tau)\;\d\sigma'=
2\pi w-\gasq P_t.
\end{eqnarray}
So it holds that
\begin{equation}\label{vpm-period}
V_\pm(z+2\pi)=\e^{\pm\i\gasq P_t}V_\pm(z),\quad
\Vb_\pm(\zb-2\pi)=\e^{\pm\i\gasq P_t}\Vb_\pm(\zb),
\end{equation}
and we can define, up to a constant normalization, new periodic
conserved quantities
\begin{equation}\label{wpm-def}
W_\pm\equiv\e^{\mp\i\gasq P_t z/(2\pi)}V_\pm,\quad
\Wb_\pm\equiv\e^{\pm\i\gasq P_t \zb/(2\pi)}\Vb_\pm.
\end{equation}
Comparing (\ref{nue-period-explizit}) and (\ref{nue-period}) the
real-valued momentum $P_t$ becomes
\begin{equation}
  \label{deltamue-explizit}
    \gasq P_t=\i\ln(ps+qr),
\end{equation}
so that the monodromy transformations (\ref{period-rand-AB}) are
restricted to those with unit determinant
\begin{equation}
  |ps+qr|=1.
\end{equation}

\section{The solution of initial value problems}
\label{sect-initial-value}

It is advantageous to use in the following calculations Kruskal
coordinates
\begin{equation}
  \label{udef}
  u=\shr{}\e^{\i t},\quad \bar u=\shr{}\e^{-\i t}.
\end{equation}
The general solution (\ref{solution}) can then be parameterized most
symmetrically by the solution of the Gelfand-Dikii equations $y_k(z)$,
$\yb_k(\zb)$
\begin{equation}\label{loes-y12}
u=\frac{\bar y_1 y_1'+\bar y_2 y_2'}{y_1 y_2'-y_1'y_2},\quad
\bar u=\frac{y_1\bar y_1'+y_2\bar y_2'}{\bar y_1\bar y_2'-\bar y_1'\bar y_2}.
\end{equation}
We shall restrict ourselves, furthermore, to regular solutions 
\begin{equation}\label{regulaer1}
y_1 y_2'-y_1'y_2\neq0,\quad\yb_1 \yb_2'-\yb_1'\yb_2\neq0\quad\forall
z,\zb,
\end{equation}
which guarantee that the Gelfand-Dikii equations are not singular,
because their coefficients, determined by
\begin{equation}
  \label{V-y}
  T=\frac{1}{\gamma^2}
  \frac{y_1''y_2'-y_1'y_2''}{y_1y_2'-y_1'y_2},\quad
  V_-=\frac{\e^{-\i\nu_0}}{\gamma^2}\left(y_1y_2'-y_1'y_2\right),
\end{equation}
are non-singular. The \glc{} invariance of the general
solution (\ref{loes-y12}) now takes the form
\begin{eqnarray}\label{GL2C-y12}
\left(\begin{array}{c}y_2\\y_1\end{array}\right)
\to
\left(\begin{array}{cc}a&-b\\c&d\end{array}\right)
\left(\begin{array}{c}y_2\\y_1\end{array}\right),\quad
\left(\begin{array}{c}\bar y_2\\\bar y_1\end{array}\right)
\to
\left(\begin{array}{cc}d&-c\\b&a\end{array}\right)
\left(\begin{array}{c}\bar y_2\\\bar y_1\end{array}\right),\\
\left(\begin{array}{cc}a&-b\\c&d\end{array}\right)\in
{\rm GL}(2,\Cset).\nonumber
\end{eqnarray}
It determines the solutions $y_k(z)$, $\yb_k(\zb)$ in terms of the
physical fields $u$, $\bar u$, just as before, up to four
indeterminate integration constants, provided we have chosen for the
physical fields the initial values at `time' $\tau_0$
\begin{equation}\label{Anfangsbed}
\!u(\sigma,\tau_0)=u_0(\sigma),\;\ub(\sigma,\tau_0)=\ub_0(\sigma),\;
\dot u(\sigma,\tau_0)=u_1(\sigma),\;
\dot \ub(\sigma,\tau_0)=\ub_1(\sigma).
\end{equation}
But in contrast to that, the chiral and anti-chiral second order
Gelfand-Dikii differential equations allow together eight integration
constants for their four independent solutions. This puzzle can be
solved as follows: differentiating the general solution
(\ref{loes-y12}), four first order differential equations result
\begin{eqnarray}\label{dgl-erster-ord}
&&y_1'=\frac{\delz\bar u}{1+u\bar u}\left(uy_1-\bar y_2\right),\quad
y_2'=\frac{\delz\bar u}{1+u\bar u}\left(uy_2+\bar y_1\right),\nonumber\\
&&\bar y_1'=\frac{\delzb u}{1+u\bar u}
\left(\bar u\bar y_1-y_2\right),\quad
\bar y_2'=\frac{\delzb u}{1+u\bar u}
\left(\bar u\bar y_2+y_1\right).
\end{eqnarray}
The elimination of the anti-chiral functions $\yb_k(\zb)$ yield,
again, the Gelfand-Dikii equations
\begin{equation}\label{Gelfand-Dikii}
  y_k''-(\delz V_-/V_-)y_k'-\gamma^2Ty_k=0,
\end{equation}
and correspondingly the anti-chiral equations result. In case, we look
now for solutions of these equations which fullfill besides the
initial state conditions (\ref{Anfangsbed}) the linear differential
equations (\ref{dgl-erster-ord}) too, the number of integration
constants is reduced from eight to the four of the \glc{} invariance
group. Fixing this \glc{} invariance the parameter functions $y_k$,
$\yb_k$ are determined, in principle, from the physical fields $u$,
$\bar u$ uniquely.

Since the coefficients of the Gelfand-Dikii equations
(\ref{Gelfand-Dikii}) are periodic functions of $z$, apart from
$y_k(z)$ also the functions $y_k(z+2\pi)$ are solutions of these
equations. They are given by (\ref{period-rand-AB}) as linear
combinations of the $y_k(z)$ 
\begin{eqnarray}\label{y-rand}
\left(
\begin{array}{c}
y_2(z+2\pi)\\y_1(z+2\pi)
\end{array}
\right)=
M
\left(
\begin{array}{c}
y_2(z)\\y_1(z)
\end{array}
\right),&&\quad
M=\matrix{cc}{p&-q\\r&s}\in\mbox{\glc}.
\end{eqnarray}
Under the \glc{} transformation $N$, the monodromy transformation $M$
changes according to
\begin{equation}
  \label{M-trafo}
  M\to NMN\inv.
\end{equation}
In case that 
\begin{equation}
  \label{glc-regulaer}
  \mbox{(1)}\quad(\tr M)^2\neq 4\det M\quad\mbox{or}\quad
  \mbox{(2)}\quad M=a\Id2,
\end{equation}
$M$ can be brought to diagonal form
\begin{equation}
  \label{M-diag}
  M=\matrix{cc}{\e^{-\alphabp}&0\\ 0&\e^{\alphap}},\quad
  \Mb=\matrix{cc}{\e^{-\alphap}&0\\ 0&\e^{\alphabp}},
\end{equation}
(the non-diag\-o\-nalizable cases can be obtained by a limiting procedure).

The periodicity conditions (\ref{y-rand}) then simplify to
\begin{eqnarray}
  \label{y-rand-spez}
  y_1(z+2\pi)=\e^\alphap y_1(z),&&\quad
  y_2(z+2\pi)=\e^{-\alphabp} y_2(z),\nonumber\\
  \yb_1(\zb-2\pi)=\e^{-\alphabp} \yb_1(\zb),&&\quad
  \yb_2(\zb-2\pi)=\e^\alphap \yb_2(\zb).
\end{eqnarray}
These properties already restrict the possible \glc{} transformations
(\ref{GL2C-y12}) to scalings with only two free parameters $a$ and $d$
\begin{equation}
  \label{y-skal}
  \matrix{c}{y_2\\ y_1}\to\matrix{c}{ay_2\\ dy_1},\quad
  \matrix{c}{\yb_2\\ \yb_1}\to\matrix{c}{d\yb_2\\ a\yb_1}.
\end{equation}
This means, we have implicitly fixed two of the four integration
constants. Any function invariant under the scalings (\ref{y-skal})
can now be determined u\-nique\-ly by (\ref{y-rand-spez}), in
particular, the periodic quotients
\begin{equation}
  \label{eta-def}
  \eta_k(z)\equiv\frac{y_k'(z)}{y_k(z)}, \quad k=1,2.
\end{equation}
They determine the solution of
(\ref{Gelfand-Dikii}) by
\begin{eqnarray}
  \label{ln-y-final}
  %y1
  \ln y_1(z)&=&\halb\intl_0^{2\pi}
  \eta_1(z')h(z-z')\d z'+
  \frac{z}{2\pi}\alpha_1+\frac{\psib_Q}{2},\nonumber\\
  %y2
  \ln y_2(z)&=&\halb\intl_0^{2\pi}
  \eta_2(z')h(z-z')\d z'+
  \frac{z}{2\pi}\alpha_2+Q_\lambda-
  \frac{\psi_Q}{2}.
\end{eqnarray}
$Q_\lambda$, $\psi_Q$ are integration constants defined by the eqs.\ 
(\ref{psi_Q-Q_lambda}) of Appendix \ref{AWP}, and $h(z)$ denotes the
periodic saw-tooth function
\begin{equation}
  \label{saege-zahn}
  h(z)=\epsilon_{2\pi}(z)-\frac{z}{\pi}=
  2n+1-\frac{z}{\pi}\quad\mbox{for }2\pi n<z<2\pi(n+1),
  \quad n\in\Zset.
\end{equation}
Here $\epsilon_{2\pi}(z)$ is the stair-step function
\begin{equation}
  \label{eps-def-kreis}
  \epsilon_{2\pi}(z)=2n+1\quad\mbox{for }2\pi n<z<2\pi(n+1),
  \quad n\in\Zset,
\end{equation}
and
\begin{equation}
  \label{alpha-12-def}
  \alpha_k=\intl_0^{2\pi}\eta_k(z)\d z
\end{equation}
are the zero modes of the fields $\eta_k(z)$.  (In passing we mention
that (\ref{y-rand-spez}), (\ref{ln-y-final}) imply
$\e^{\alpha'}=\e^{\alpha_1}$, and we define $\alpha'\equiv\alpha_1$.)
But we have to stress here especially that this result does not
deliver $\eta_k$ or $y_k$ explicitly as functions of $u$, $\bar u$.
However, it will be sufficient in order to calculate their Poisson
brackets.

\section{The Poisson brackets}
\label{sect-pk-y-kreis}

We calculate Poisson brackets by assuming canonical Poisson brackets
of the physical fields, which are obtained from the action
(\ref{action}). For the Kruskal coordinates we get the following
non-vanishing expressions
\begin{eqnarray}\label{pk-uuq}
&&\{u(\sigma,\tau),\dot{\bar u}(\sigma',\tau)\}=
\{\bar u(\sigma,\tau),\dot{u}(\sigma',\tau)\}=
2\gamma^2(1+u\bar u)\deltaper(\sigma-\sigma'),\nonumber\\
&&\{\dot u(\sigma),\dot{\bar u}(\sigma')\}=
2\gamma^2(\dot u\bar u-
u\dot{\bar u})\deltaper(\sigma-\sigma'),
\end{eqnarray}
where $\deltaper$ is the periodic $\delta$-function defined by
\begin{equation}\label{delta-kreis}
  \delta_{2\pi}(\sigma-\sigma')\equiv
  \sum_{n=-\infty}^\infty\delta(\sigma-\sigma'+2\pi n).
\end{equation}

This allows us to calculate Poisson brackets of all quantities
explicitly expressed in terms of the physical fields. Here we want to
determine the Poisson brackets of the parameter functions $y_k(z)$.
We saw in the preceding section, that the $\eta_k$ are uniquely
defined by the initial state conditions, but they were not given
explicitly as functions of the physical fields. We shall show that the
Poisson brackets of the $\eta_k$, and the $y_k$ can, nevertheless, be
derived.

First, we have to determine the Poisson brackets of the $\eta_k(z)$,
and those of the $y_k$ then follow by means of (\ref{ln-y-final}). We
calculate the variations $\delta\eta_k(z)$ as functions of the varied
physical fields and momenta by varying the Gelfand-Dikii equations
\begin{equation}\label{var-y-gl}
  \delta y_k''-(\delz V_-/V_-)\delta y_k'-\gamma^2T\delta y_k=
  \delta (\delz V_-/V_-)y_k'+\gamma^2\delta Ty_k.
\end{equation}
We vary, as well, the boundary conditions (\ref{y-rand-spez}),
eliminate $\delta\alphap$ and $\delta\alphabp$, and get subsidiary
conditions for the equations (\ref{var-y-gl})
\begin{eqnarray}
  \label{dy-rand-kreis}
  \hspace*{-20pt}
  y_k'(z)\dey_k(z+2\pi)-y_k(z)\dey_k'(z+2\pi)&=&
  y_k'(z+2\pi)\dey_k(z)-y_k(z+2\pi)\dey_k'(z).\nonumber\\
  \end{eqnarray}
  They restrict the general solution of (\ref{var-y-gl}), which is
  defined by a special solution of (\ref{var-y-gl}) and the general
  solution of (\ref{Gelfand-Dikii}), to
\begin{eqnarray}
  \label{var-y-sol-kreis}
  \hspace*{-13pt}
  \dey_k(z)&=&\intk\!\Omega_k(z,z')
  \left(\delta(\partial V_-/V_-)(z')y_k'(z')+
  \gamma^2\delta T(z')y_k(z')\right)\d z'+\delta C_ky_k(z),\nonumber \\
  &&\Omega_1(z,z')\equiv
  \frac{y_2(z')y_1(z)}{y_1(z')y_2'(z')-y_2(z')y_1'(z')}
  \frac{E(z,z')-\epsper(z-z')}{2},\nonumber\\
  &&\Omega_2(z,z')\equiv
  \frac{y_1(z')y_2(z)}{y_1(z')y_2'(z')-y_2(z')y_1'(z')}
  \frac{E(z',z)+\epsper(z-z')}{2},
  \nonumber\\
  &&E(z,z')\equiv
  \frac{\displaystyle\exp\left(
    \frac{\alpha_1-\alpha_2}{2}\epsper(z-z')\right)}
  {\displaystyle\sinh\frac{\alpha_1-\alpha_2}{2}}
  \frac{y_2(z)y_1(z')}{y_1(z)y_2(z')}.
\end{eqnarray}
The variations $\deC_k$ correspond to the undetermined scalings
(\ref{y-skal}). They cannot simply be set to zero because we are
integrating (\ref{var-y-sol-kreis}) over non-periodic functions of
$z'$, and without the term $\deC_ky_k(z)$ these integrals would depend on a
shift of the integration range.

Since the functions $\eta_k(z)$ do not depend on the scalings
(\ref{y-skal}), their variations 
\begin{equation}
  \label{var-eta}
  \delta\eta_k(z)=\delta\frac{y_k'(z)}{y_k(z)}=
  \frac{\dey_k'(z)}{y_k(z)}-
  \frac{y_k'(z)\dey_k(z)}{y_k(z)^2}
\end{equation}
show, indeed, the cancelation of $\delta C_k$ in
\begin{eqnarray}
  \label{var-eta-sol}
  \delta\eta_k(z)&=&\intk\omega_k(z,z')
  \left(\delta(\partial V_-/V_-)(z')\eta_k'(z')+
  \gamma^2\delta T(z')\eta_k(z')\right)\d z',\nonumber\\
  &&\omega_1(z,z')\equiv\halb E(z,z')
  \frac{\eta_1(z)-\eta_2(z)}{\eta_1(z')-\eta_2(z')},\nonumber\\  
  &&\omega_2(z,z')\equiv-\halb E(z',z)
  \frac{\eta_1(z)-\eta_2(z)}{\eta_1(z')-\eta_2(z')}.
\end{eqnarray}
The integrands of (\ref{var-eta-sol}) are periodic functions of
the integration variable, and (\ref{var-eta-sol}) determines the
Poisson brackets of $\eta_k(z)$. The non-vanishing ones are given by
\begin{eqnarray}
  \label{pk-eta}
  \{\eta_1(z),\eta_2(z')\}&=&\gasqhalb
  (\eta_1(z)-\eta_2(z))E(z,z')(\eta_1(z')-\eta_2(z')) \nonumber\\
  &&-\gasq(\eta_1(z)-\eta_2(z))\deltaper(z-z').
\end{eqnarray}
The Poisson brackets of the functions $\ln y_k$  result, finally, by
means of (\ref{ln-y-final})
\begin{eqnarray}
  \label{pk-y-kreis}
  \pk{\ln y_1(z)}{\ln y_1(z')}&=&0,\nonumber\\
  \pk{\ln y_1(z)}{\ln y_2(z')}&=&
  \gasqhalb\left(\epsper(z-z')-\frac{z-z'}{2\pi}\right)-
  \gasqhalb E(z,z')+\nonumber\\
  &&{}+\gasqapi\intk\d z E(z,z'),\nonumber\\
  \pk{\ln y_1(z)}{\ln\yb_1(\zb')}&=&-\gasqvpi (z-\zb'),\nonumber\\
  \pk{\ln\yb_1(\zb)}{\ln y_2(z')}&=&-\gasqapi\intk\d z 
  \frac{\eta_2(z)}{\eta_1(z)}E(z,z'),\\
  \pk{\ln y_2(z)}{\ln y_2(z')}&=&
  -\gasqapi\intk\d z\frac{\eta_2(z)}{\eta_1(z)}E(z,z')+
  \gasqapi\intk\d z'\frac{\eta_2(z')}{\eta_1(z')}E(z',z),
  \nonumber\\
  \pk{\ln y_2(z)}{\ln\yb_2(\zb')}&=&-\gasqvpi(z-\zb')-
  \gasqapi\intk\!\!\d z' E(z',z)+
  \gasqapi\intk\!\!\d\zb\Eb(\zb,\zb').\nonumber
\end{eqnarray}
In distinction to the field theoretic results of ref. \cite{cmp} we
observe here a structurally changed non-local realization of the
algebra. This is due to zero modes which arise additionally in the
periodic case. It might be surprising that the algebra treats $y_1$
and $y_2$ (as well as $\yb_1$ and $\yb_2$) asymmetrically. A symmetric
treatment of the functions $y_k$ and $\yb_k$ is presented in appendix
\ref{anhang-symm-pk}. But it turns out that the algebra
(\ref{pk-y-kreis}) is more appropriate for transforming the $y_k$,
$\yb_k$ onto canonical free fields.

\section{The canonical transformation onto periodic free fields}
\label{frei-trafo-kreis}

There are several methods to find relations between $y_k(z)$, $\bar
y_k(\zb)$ and chiral, respectively anti-chiral components $\phi_k(z)$,
$\phib_k(\zb)$ of canonical free fields ($k=1,2$)
\begin{equation}\label{psi12-komp}
\psi_k(\sigma,\tau)=\phi_k(z)+\phib_k(\zb).
\end{equation}
Sometimes it will be useful to have  the mode expansions in mind, e.\ g., 
\begin{equation}
  \phi_k(z)=\frac{q_k}{2}+\left(\frac{p_k}{4\pi}+\frac{w'}{2\gamma}
  \delta_{k,2}\right)z+\frac{\i}{\sqrt{4\pi}}\sum_{n\neq0}
  \frac{a_n^{(k)}}{n}\e^{-\i nz}.
\end{equation}
$w'$ is an integer `winding number'.

Here we simply assume that there is a free-field theory with a
corresponding free-field energy-mo\-men\-tum tensor which is
canonically related to our \sll{} model. Then the easiest and most
straightforward approach identifies the energy-mo\-men\-tum tensors
of both theories
\begin{equation}\label{T-phi-y}
T(z)=\left(\delz\phi_1\right)^2+\left(\delz\phi_2\right)^2
=\frac{1}{\gamma^2}
\frac{y_1''y_2'-y_1'y_2''}{y_1y_2'-y_1'y_2}.
\end{equation}
Furthermore, we assume that the free fields $\psi_1$, $\psi_2$ are
local expressions of the parameter functions $y_k$.

It is appropriate to introduce complex free fields
\begin{equation}\label{psi-komplex}
\psi=\psi_1+\i\psi_2,\quad\psib=\psi_1-\i\psi_2,
\end{equation}
which factorize the components of the energy-momentum tensor
\begin{equation}\label{T-frei-komp}
T(z)=\delz\psi\delz\psib,\quad
\bar T(z)=\delzb\psi\delzb\psib.
\end{equation}

(\ref{psi12-komp}) gives a corresponding chiral decomposition of
$\psi$ and $\bar{\psi}$
\begin{equation}\label{psi-komp}
\psi(\sigma,\tau)=\phi(z)+\chib(\zb),\quad
\psib(\sigma,\tau)=\chi(z)+\phib(\zb),
\end{equation}
with
\begin{eqnarray}\label{phi-def}
\phi(z)=\phi_1(z)+\i\phi_2(z),&&\quad
\phib(\zb)=\phib_1(\zb)-\i\phib_2(\zb),\nonumber\\
\chi(z)=\phi_1(z)-\i\phi_2(z),&&\quad
\chib(\zb)=\phib_1(\zb)+\i\phib_2(\zb).
\end{eqnarray}

The most general solution of this problem depends on several complex
constants, and it is given by \cite{Mu}
\begin{eqnarray}\label{phi-y}
\phi=\frac{1}{\gamma C}\left(
\ln\frac{\alpha y_1'+\beta y_2'}{y_1 y_2'-y_1'y_2}+D\right),
&&\quad\chi=\frac{C}{\gamma}\ln\left(\alpha y_1+\beta y_2\right),
\nonumber\\
\phib=\frac{1}{\gamma\Cb}\left(
\ln\frac{\bar\alpha \yb_1'+\bar\beta \yb_2'}{\yb_1 \yb_2'-\yb_1'\yb_2}+
\Db\right),
&&\quad\chib=\frac{\bar C}{\gamma}
\ln\left(\bar\alpha \yb_1+\bar\beta \yb_2\right).
\end{eqnarray}
But the constants can be further restricted.  Taking into
consideration the invariance of (\ref{T-frei-komp}) under
$\phi\to\e^{\i\delta}\phi$, $\chi\to\e^{-\i\delta}\chi$, we can choose
$C$ real positive.  Of course, the physics should not depend on the
choice of the branch of the logarithm. This implies $C=1$.
Furthermore, $\phi_2$ is defined modulo $2\pi/\gamma$ only (i.e.\ 
$\phi_2$ takes values on a circle with radius $1/\gamma$)
\begin{equation}
  \label{phi2-equiv}
  \phi_2\equiv\phi_2+\frac{2\pi}{\gamma},\quad
  \phib_2\equiv\phib_2+\frac{2\pi}{\gamma}.
\end{equation}
Up to winding contributions the free fields $\psi_k$ are assumed
periodic
\begin{equation}
  \label{psi-monodr}
  \psi_1(\sigma+2\pi,\tau)=\psi_1(\sigma,\tau),\;\;
  \psi_2(\sigma+2\pi,\tau)=\psi_2(\sigma,\tau)+\frac{2\pi w'}{\gamma},
  \;\; w'\in\Zset,
\end{equation}
and we obtain
\begin{eqnarray}
  \label{phi-monodr}
  \phi_k(z+2\pi)-\phi(z)&=&\frac{p_k}{2} +
  \delta_{k,2}\frac{\pi w'}{2\gamma},\nonumber\\
  \phib_k(\zb+2\pi)-\phib(\zb)&=&\frac{p_k}{2} -
  \delta_{k,2}\frac{\pi w'}{2\gamma},
  \quad p_k\in\Rset.
\end{eqnarray}
But this is consistent with (\ref{phi-y}) and (\ref{y-rand-spez}) only
if one of the pairs $(\alpha,\alphab)$ and $(\beta,\betab)$ is
$(0,0)$. Choosing $\beta=\betab=0$, the rescaling (\ref{y-skal}) now
allows us to set $\alpha=\alphab=1$, and considering the invariance under
$\psi_k\to\psi_k+\const$, we can without loss of generality fix $D=\Db=0$. 
Thus the solution (\ref{phi-y}) simplifies finally to
\begin{eqnarray}
  \label{phi-y-spez-kreis}
  \phi=\frac{1}{\gamma} 
  \ln\frac{y_1'}{y_1 y_2'-y_1'y_2},\quad
  &&\chi=\frac{1}{\gamma}\ln y_1, \nonumber\\ 
  \phib=\frac{1}{\gamma}
  \ln\frac{\yb_1'}{\yb_1\yb_2'-\yb_1'\yb_2},\quad
  &&\chib=\frac{1}{\gamma}\ln\yb_1.
\end{eqnarray}
As expected, from the non-local Poisson bracket relations
(\ref{pk-y-kreis}) we get for the fields $\phi_k$, $\phib_k$, indeed,
the local free-field Poisson brackets
\begin{eqnarray}
  \label{pk-frei-kreis}
  \pk{\phi_k(\tau+\sigma)}{\phi_l(\tau+\sigma')}&=&
  -\frac{\delta_{kl}}{4}\left(\epsper(\sigma-\sigma')-
  \frac{\sigma-\sigma'}{2\pi}\right),
  \nonumber\\
  \pk{\phib_k(\tau-\sigma)}{\phib_l(\tau-\sigma')}&=&
  \frac{\delta_{kl}}{4}\left(\epsper(\sigma-\sigma')-
  \frac{\sigma-\sigma'}{2\pi}\right),
  \nonumber\\
  \pk{\phi_k(\tau+\sigma)}{\phib_l(\tau-\sigma')}&=&
  -\frac{\delta_{kl}}{8\pi}\left(\sigma+\sigma'\right).
\end{eqnarray}
Solving now (\ref{phi-y-spez-kreis}) for $y_k$, $\yb_k$, their
non-local free-field representation result
\begin{eqnarray}
  \label{y-phi-kreis}
  y_1(z)&=&\exp\left(\gamma\chi(z)\right),\nonumber\\
  y_2(z)&=&-\frac{\exp\left(\gamma\chi(z)\right)}
  {2\sinh(\gamma p_1/2)}
  \intk\d z'\;\gamma\chi'(z')
  \exp\left(-\frac{\gamma p_1}{2}\epsper(z-z')-
  2\gamma\phi_1(z')\right),\nonumber\\
  \yb_1(\zb)&=&\exp\left(\gamma\chib(\zb)\right),\\
  \yb_2(\zb)&=&-\frac{\exp\left(\gamma\chib(\zb)\right)}
  {2\sinh(\gamma p_1/2)}
  \intk\d\zb'\;\gamma\chib'(\zb')
  \exp\left(-\frac{\gamma p_1}{2}\epsper(\zb-\zb')-
  2\gamma\phib_1(\zb')\right),\nonumber
\end{eqnarray}
where the zero mode momentum is given by
\begin{equation}
  \label{p1-def}
  p_1=\intl_0^{2\pi}\dot\psi_1(\tau,\sigma)\d\sigma.
\end{equation}

We have checked that the free-field Poisson brackets
(\ref{pk-frei-kreis}) yield, conversely, the non-local Poisson brackets
of the $y_k(z)$, $\bar y_k(\zb)$ (\ref{pk-y-kreis}), and we could show
that these results also follow from the Gelfand-Dikii equations
(\ref{Gelfand-Dikii}), in case, their coefficients are expressed in
terms of the free fields and the initial state problem is solved anew.

This proves that these free-field transformations of
the physical fields $r$, $t$, or $u$, $\ub$ are canonical
transformations, and they are one to one.

\section{Summary}
We have completely integrated the periodic \slu{} gauged WZNW theory and
calculated its symplectic structure. This allows us to relate this model
canonically to a free-field theory. The results could be
summarized in terms of local B\"acklund transformations which are
identical to those of the non-periodic case \cite{cmp}. Instead, we
give here the complete canonical transformation of the fields
$u(\sigma,\tau)$, $\ub(\sigma,\tau)$ onto the free fields
\begin{eqnarray}
  u&=&\e^{\gamma(\phi+\chib)}\left(1+\Phi\Phib\right)-
  \frac{1}{4}\e^{-\gamma(\phib+\chi)}
  +\frac{\i}{2}\left(\e^{\gamma(\phi-\phib)}\Phi+
    \e^{-\gamma(\chi-\chib)}\Phib\right),\nonumber\\
  \ub&=&\e^{\gamma(\phib+\chi)}\left(1+\Phi\Phib\right)-
  \frac{1}{4}\e^{-\gamma(\phi+\chib)}
  -\frac{\i}{2}\left(\e^{\gamma(\chi-\chib)}\Phi+
    \e^{-\gamma(\phi-\phib)}\Phib\right).
\end{eqnarray}
This transformation is non-locally defined by
\begin{eqnarray}
  \Phi(z)&=&-\frac{1}{2\sinh(\gamma p_1/2)}
  \int\limits_0^{2\pi}\!\!
  \d z'\;\gamma\phi_2'(z')\exp\left(
    -\frac{\gamma p_1}{2}\epsilon_{2\pi}(z-z')-
    2\gamma\phi_1(z')\right),\nonumber\\[-2mm]
  \\[-2mm]
  \Phib(\zb)&=&-\frac{1}{2\sinh(\gamma p_1/2)}
  \int\limits_0^{2\pi}\!\!
  \d\zb'\;\gamma\phib_2'(\zb')
  \exp\left(-\frac{\gamma p_1}{2}\epsilon_{2\pi}(\zb-\zb')-
    2\gamma\phib_1(\zb')\right).\nonumber
\end{eqnarray}
As in Liouville theory this structure might require quantum mechanical
deformations. But we might be confronted, as well, with further
unusual problems related to the quantization of the parafermionic
structure of the theory, which is classically defined by non-linear
Poisson brackets. Using the freedom of normalization of
(\ref{wpm-def}) (here we take into consideration the full $q$ zero
mode of the free field) for the periodic case the parafermions
fulfil
\begin{eqnarray}\label{valgebra}
  \{W_\pm(z),W_\pm(z')\}&=&\gamma^2
  W_\pm(z)W_\pm(z')\,h(z-z'),\nonumber\\ 
  \{W_\pm(z),W_\mp(z')\}&=&-\gamma^2
  W_\pm(z)W_\mp(z')\,h(z-z')+
  \frac{1}{\gamma^2}\left(\delz 
    +\frac{i \gamma p_2}{2\pi}\right)\delta_{2\pi}(z-z'),\nonumber\\ 
  \{p_2,W_\pm(z')\}&=&\mp2i\gamma W_\pm(z'), 
\end{eqnarray}
In our opinion this may be a good starting point for a quantization.
Therefore, it remains a challange to implement the exact canonical
quantization of the \slu{} model on the basis of our results.

\section*{Acknowledgement}
We would very much like to thank C. Ford and G. Jorjadze for
reading the manuscript and for useful discussions.

\begin{appendix}

\section{The definition of the integration constants}
\label{AWP}
Here we explain the relations (\ref{ln-y-final}) in more detail.

Using chirality, the functions $\eta_k(z)$ of (\ref{eta-def}) can be
integrated to
\begin{equation}
  \label{ln-y-int}
  \ln y_k(z)=\halb\intl_0^{2\pi}
  \eta_k(z')h(z-z')\d z'+
  \frac{z}{2\pi}\alpha_k+D_k.
\end{equation}
$h(z)$ and $\alpha_k$ are defined by (\ref{saege-zahn}),
(\ref{alpha-12-def}). $D_k$ are the integration constants under
discussion. Let us consider the canonical free field
(\ref{psi-komplex}), (\ref{phi-y-spez-kreis})
\begin{equation}\label{psi-y-spez}
\psi(\sigma,\tau)=\frac{1}{\gamma}
\ln\frac{y_1'(z)}{y_1(z) y_2'(z)-y_1'(z)y_2(z)}+\frac{1}{\gamma}
\ln\yb_1(\zb).
\end{equation}
Equations (\ref{eta-def}) and (\ref{dgl-erster-ord}) allow one to replace
the functions $y_k'(z)$, $y_k(z)$ and $\yb_1(\zb)$ by $\eta_k(z)$.
Then, $\psi$ is completely given in terms of $\eta_k(z)$ and the
physical fields $u$, $\ub$
\begin{equation}\label{psi-y-spez1}
\psi(\sigma,\tau)=\frac{1}{\gamma}
\ln\left(\frac{u\delz\ub-\eta_2(1+u\ub)}
  {\delz\ub(\eta_1-\eta_2)}\eta_1\right).
\end{equation}
{}Similarly, eqs.\ (\ref{dgl-erster-ord}) yield
\begin{equation}
  \label{eta-etab-zushang}
  \eta_{1,2}=\frac{\delzb u-\etab_{2,1}u}
  {\ub\delzb u-\etab_{2,1}(1+u\ub)}\delz\ub,
\end{equation}
which shows that we could express $\psi$, as well, in terms of
$\etab_k(\zb)$ and $u$, $\ub$.

On the other hand, substituting (\ref{ln-y-int}) into
(\ref{psi-y-spez}) we obtain
\begin{eqnarray}
  \label{psi-eta}
  \psi(\sigma,\tau)&=&\halb\intk\left[
  \eta_1(\tau+\sigma')+\etab_2(\tau-\sigma')\right]
  h(\sigma-\sigma')\d\sigma'+\nonumber\\
  &&+\frac{\tau}{2\pi}(\alpha_1-\alphab_2)+
  \i\sigma m+D_1-\Db_2+\lambda(z).
\end{eqnarray}
Here 
\begin{equation}
  \label{lambda-def}
  \lambda(z)\equiv\ln\eta_1(z)-\ln(\eta_2(z)-\eta_1(z)),
\end{equation}
and $m$, defined by
\begin{equation}
  \label{m-def}
  \alpha_1+\alphab_2=2\pi\i m,
\end{equation}
is an integer due to the periodicity of $y_1(z)/\yb_2(\zb)$ (cp.\ 
(\ref{y-rand-spez}) with the (anti-) chiral (\ref{ln-y-int})).
$D_1-\Db_2$ is uniquely determined by the constant zero modes of
$\psi$ and $\lambda$
\begin{equation}
  \label{D-zero}
  D_1-\Db_2=\psi_Q-Q_\lambda.
\end{equation}
Since the two parts of (\ref{psi-y-spez}) are the chiral and
anti-chiral components of $\psi$, and distributing the zero mode
$\psi_Q$ half and half to these components, by comparison we obtain
\begin{equation}
  \label{D-def}
  D_1=\frac{\psi_Q}{2}+Q_\lambda,\quad
  \Db_1=\frac{\psib_Q}{2}+\Qb_\lambda,\quad
  D_2=-\frac{\psib_Q}{2},\quad  
  \Db_2=-\frac{\psi_Q}{2}.
\end{equation}
This immediately implies (\ref{ln-y-final}). We give, finally, the
explicit expressions for the constants $\psi_Q$ and $Q_\lambda$ in
terms of the $\eta_k$
\begin{eqnarray}
  \label{psi_Q-Q_lambda}
  \psi_Q&=&\frac{1}{2\pi}\intk\left(
  \psi(\sigma,\tau)-\psi_P(\sigma,\tau)\right)\d\sigma,\mbox{ with }\nonumber\\
  \psi_P(\sigma,\tau)&=&\frac{\sigma}{2\pi}\intk
  \frac{\del\psi(\sigma',\tau)}{\del\sigma'}\d\sigma'+
  \frac{\tau}{2\pi}\intk
  \frac{\del\psi(\sigma',\tau)}{\del\tau}\d\sigma'\nonumber\\
  Q_\lambda&=&\frac{1}{2\pi}\intk
  \left(\lambda(z)-\frac{P_\lambda}{2\pi}z\right)\d z,\mbox{ with }\nonumber\\
  P_\lambda&=&\intk\lambda'(z)\d z=
  \intk\frac{\eta_1(z)\eta_2'(z)-\eta_1'(z)\eta_2(z)}
  {\eta_1(z)\left(\eta_1(z)-\eta_2(z)\right)}\d z,
\end{eqnarray}
which also determine their commutation relations.

\section{Symmetric Poisson brackets}
\label{anhang-symm-pk}

In this appendix we define functions $\ys_k(z)$ and $\ybs_k(\zb)$ with
Poisson brackets which are symmetric under the exchange 
$1\leftrightarrow 2$.

Using the shorthand notation
\begin{eqnarray}
  \label{kappa-null-const}
  \kappa_Q&=&\frac{1}{2\pi}\intk\left[\ln y_1(\tau+\sigma)-
    \ln\yb_2(\tau-\sigma)-
  \frac{\tau}{2\pi}(\alpha_1-\alphab_2)-\i\sigma m\right]\d\sigma,
  \nonumber\\
  \kappab_Q&=&\frac{1}{2\pi}\intk\left[\ln\yb_1(\tau-\sigma)-
    \ln y_2(\tau+\sigma)
  -\frac{\tau}{2\pi}(\alphab_1-\alpha_2)-\i\sigma\mb\right]\d\sigma,
\end{eqnarray}
these functions are defined by
\begin{eqnarray}
  \label{ys-y-zushang}
  \ln\ys_1(z)=
  \ln y_1(z)+\frac{\kappa_Q}{2}-\frac{\psib_Q}{2},&&
  \ln\ys_2(z)=
  \ln y_2(z)+\frac{\kappab_Q}{2}-\frac{\psi_Q}{2},\nonumber\\
  \ln\ybs_1(\zb)=
  \ln\yb_1(\zb)+\frac{\kappab_Q}{2}-\frac{\psi_Q}{2},&&
  \ln\ybs_2(\zb)=
  \ln\yb_2(\zb)+\frac{\kappa_Q}{2}-\frac{\psib_Q}{2}.
\end{eqnarray}
The symmetric Poisson brackets are
\begin{eqnarray}
  \label{pk-ys-kreis}
  \pk{\ln\ys_1(z)}{\ln\ys_2(z')}&=&
  \gasqhalb\left(\epsper(z-z')-\frac{z-z'}{2\pi}\right)-
  \gasqhalb E(z,z')+\nonumber\\
  &&{}+\gasqapi\intk\d zE(z,z')+\gasqapi\intk\d z'E(z,z')-\nonumber\\
  &&{}-\gasqzpisq\intk\intk\d z\d z'(E(z,z')-\Eb(z',z)),\nonumber\\
  &&\hspace{-3.8cm}\pk{\ys_1(z)}{\ys_1(z')}=\pk{\ys_2(z)}{\ys_2(z')}=
  \pk{\ys_1(z)}{\ybs_2(\zb')}=\pk{\ybs_1(\zb)}{\ys_2(z')}=0,\nonumber\\ 
  \pk{\ln\ys_1(z)}{\ln\ybs_1(\zb')}&=&-\frac{\gamma^2}{4\pi}(z-\zb')+
  \gasqapi\intk\d z'E(z,z')-\nonumber\\
  &&{}\hspace{-2cm}-\gasqapi\intk\d\zb\Eb(\zb',\zb)-
  \gasqzpisq\intk\intk\d z\d z'(E(z,z')-\Eb(z',z)),\nonumber\\
  \pk{\ln\ybs_2(\zb)}{\ln\ys_2(z')}&=&-\frac{\gamma^2}{4\pi}(\zb-z')+
  \gasqapi\intk\d zE(z,z')-\\
  &&{}\hspace{-2cm}-\gasqapi\intk\d\zb'\Eb(\zb',\zb)-
  \gasqzpisq\intk\intk\d z\d z'(E(z,z')-\Eb(z',z)).\nonumber
\end{eqnarray}

\end{appendix}

\end{document}